\documentclass[aps,preprint]{revtex4}
\usepackage[T1]{fontenc}
\usepackage{mathptmx}
\usepackage{siunitx}
\usepackage{soul}
\usepackage{graphicx}
\usepackage{epstopdf}
\usepackage{ulem}
\usepackage{subfigure}
\usepackage{float}
\usepackage{dcolumn}
\usepackage{bm}
\usepackage{booktabs}
\usepackage{amsmath,amsthm,mathrsfs,amsfonts,amssymb}
\usepackage{slashed}
\usepackage{cancel}
\usepackage{bpchem}
\usepackage{lipsum}
\usepackage{xcolor}
\usepackage[colorlinks,linkcolor=blue,citecolor=blue,urlcolor=blue]{hyperref}
\usepackage{cleveref}
\usepackage{listings}

\allowdisplaybreaks[4]
\DeclareMathAlphabet{\mathcal}{OMS}{cmsy}{m}{n}
\DeclareSymbolFont{Letters}{OML}{cmm}{m}{it}
\DeclareMathSymbol{\psi}{\mathalpha}{Letters}{32}
\DeclareMathSymbol{\Psi}{\mathalpha}{Letters}{9}
\crefname{equation}{Eq.}{Eqs.}
\crefname{figure}{Fig.}{Figs.}
\crefname{table}{Tab.}{Tabs.}
\crefname{section}{Sec.}{Sec.}
\newcommand{\beq}{\begin{equation}}
\newcommand{\eeq}{\end{equation}}
\newcommand{\ben}{\begin{align}}
\newcommand{\een}{\end{align}}
\newcommand{\bea}{\begin{aligned}}
\newcommand{\eea}{\end{aligned}}
\newcommand{\bes}{\begin{subequations}}
\newcommand{\ees}{\end{subequations}}
\newcommand{\bew}{\begin{widetext}}
\newcommand{\eew}{\end{widetext}}

\numberwithin{table}{section}

\begin{document}
\preprint{CTP-SCU/2020020}
\title{Joule-Thomson Expansion of Born-Infeld AdS Black Holes}
\author{Shihao Bi $^{a}$}
\email{bishihao@stu.scu.edu.cn}
\author{Minghao Du$^{a}$}
\email{duminghao@stu.scu.edu.cn}
\author{Jun Tao$^{a}$}
\email{taojun@scu.edu.cn}
\author{Feiyu Yao$^{a}$}
\email{yaofeiyu@stu.scu.edu.cn}
\affiliation{$^{a}$Center for Theoretical Physics, College of Physics, Sichuan University, Chengdu, 610065, China}

\begin{abstract}
In this paper, the Joule-Thomson expansion of Born-Infeld AdS black holes is studied in the extended phase space, 
where the cosmological constant is identified with the pressure. The Joule-Thomson coefficient, 
the inversion curves and the isenthalpic curves are discussed in detail by 4-dimensional black hole. 
The critical point of Born-Infeld black hole is depicted with varying parameter $\beta$ and the charge $Q$. 
In $T-P$ plane, the inversion temperature curves and isenthalpic curves are obtained with different parameter $\beta$ and the charge $Q$. 
We find that the missing negative slope is still conserved in Born-Infeld black holes. 
We also extend our discussion to arbitrary dimension higher than 4.
The critical temperature and the minimum of inversion temperature are compared, and the ratio is asymptotically $1/2$ as $Q$ increases or $\beta\to\infty$ in $D=4$,
and reproduce the previous results in higher dimension.
\end{abstract}

\maketitle

\section{Introduction}

The pioneer work \cite{Bekenstein:1974ax,Hawking:1974rv} of Hawking and Bekenstein makes it possible to study black holes from the perspective of thermodynamics. Since the four laws of black hole mechanics were established \cite{Bardeen:1973gs}, much attention has been paid to reveal the deep and fundamental relationships between the laws of general relativity, quantum field theory and thermodynamics \cite{Klebanov:1999tb,Hawking:1976de,Wald:1999vt,Witten:1998zw}. In view of the key role played in the research of quantum gravity \cite{Strominger:1996sh}, surrounding the topic of black holes many ideas collide and new insights have extensively enriched our understanding of the universe, such as the holographic superfluids and superconductors \cite{Strominger:1996sh,Hartnoll:2008vx,Herzog:2008he,Herzog:2011ec}, the quantum version in condensed matter theory \cite{Kitaev,Volovik:2003fe,Maldacena:2016hyu,Giovanazzi:2004zv}, etc.

Hawking and Page showed the phase transition between the Schwarzschild AdS black hole and the thermal AdS space\cite{Hawking:1982dh}. It has been shown that black holes in AdS space share common properties with general thermodynamic systems, and this relation was further enhanced in the extended phase space\cite{Kubiznak:2012wp}, where the cosmological constant and its conjugate quantity are treated as the thermodynamic pressure and volume, respectively 
\begin{equation}
P=-\frac{\Lambda }{8\pi }=\frac{(D-1)(D-2)}{16 \pi l^2},\ \ \  V=\left( \frac{\partial M}{\partial P}\right)
_{S,Q},  \label{eq:pv}
\end{equation}
where $l$ is the AdS space radius and the black hole mass $M$ is understood as the enthalpy \cite{Kubiznak:2016qmn}. Then a number of papers explored   various thermodynamic aspects of black holes such as phase transition \cite{Altamirano:2013uqa,Wei:2014hba}, heat engine's efficiency \cite{Johnson:2014yja,Hendi:2017bys}, compressibility \cite{Dolan:2011jm,Dolan:2013dga}, critical phenomenon \cite{Wei:2012ui,Banerjee:2011cz,Niu:2011tb}, weak cosmic censorship conjecture\cite{pen,Chen:2019pdj}. A brief review is given in ref. \cite{Kubiznak:2016qmn}.

Non-linear electrodynamics (NLED) is the natural extension of the Maxwell's theory in extreme conditions such as high field and small length scale. 
In recent years, nonlinear electrodynamics has been widely investigated in the context of black hole physics, for its promising construction
of regular black hole solutions. Among the NLED theories, the Born-Infeld electrodynamics has attracted a lot of attentions and aroused popular research interest.
The Born-Infeld electrodynamics encodes the low-energy dynamics of D-branes, which incorporates maximal electric fields and smooths out the divergence of the electrostatic self-energy of point-like charges. The Born-Infeld AdS black hole solution was first obtained in \cite{Cai:2004eh}. The 
thermodynamic behaviors and phase transitions of these black holes were studied both in the canonical \cite{Fernando:2003tz} and the grand canonical 
ensembles \cite{Fernando:2006gh}. Moreover, we can consider the string corrections to the thermodynamic properties of charged AdS black holes by introducing the Born-Infeld action  \cite{Born:1933qff,Born:1934ji,Tao:2017fsy,Liang:2019dni}. Due to its intimate connection with string theory, a candidate theory of quantum gravity, we believe such an extension would bring some new insight when it comes to the quantum regime.
 
The Joule-Thomson expansion of black holes was first investigated in \cite{Okcu:2016tgt}. This subject subsequently was comprehensively studied in \cite{Okcu:2017qgo,Mo:2018rgq,Lan:2018nnp,Chabab:2018zix,Ghaffarnejad:2018exz,AhmedRizwan:2019yxk,Pu:2019bxf,Li:2019jcd,Mo:2018qkt,Cisterna:2018jqg,Haldar:2018cks,Yekta:2019wmt,Rostami:2019ivr,Kuang:2018goo,Guo:2020qxy,Guo:2020mpla,K.:2020rzl,Hegde:2020xlv,Nam:2020epjp,Sadeghi:2020npb,Lan:2018npb,Zhao:2018prd}.  The inversion curves that separate heating-cooling regions in $T-P$ plane for isenthalpic curves with different parameters were presented. The results of these papers show that the inversion curves $T(P)$ for different black hole systems are similar. In this paper, we would like to generalize the current research of Joule-Thomson expansion to the case of  the Born-Infeld AdS black holes. 

This paper is organized as follows. In \cref{sec:bht} we briefly review the thermodynamics of  the Born-Infeld AdS black holes in $D$-dimensional spacetime. Then in \cref{sec:jte} we discuss the Joule-Thomson expansion of the Born-Infeld AdS black hole in $D$-dimensional spacetime, including the Joule-Thomson coefficient, the inversion curves and the isenthalpic curves. Furthermore, we compare the critical temperature and the minimum of inversion temperature. The influence of the nonlinear electrodynamics parameters and the charge $Q$ on inversion curves is also discussed. 
Finally we make a conclusion in \cref{sec:con}. We will use the units $G=\hslash=k_{B}=c=1$.

\section{Black Hole Thermodynamics}
\label{sec:bht}
The Einstein-Hilbert-Born-Infeld action for $D$-dimensional ($D\geqslant 4$) spacetime could be described as follows 
\begin{equation}
S=\frac{1}{16\pi }\int \mathrm{d}^{D}x\sqrt{-g}(R-2\Lambda +\mathcal{L}(F)),
\end{equation}
where the cosmological constant is related to the AdS space radius $l$ by 
\begin{equation}
\Lambda =-\frac{(D-1)(D-2)}{2l^{2}}.  \label{eq:cc}
\end{equation}
This action is a nonlinear generalization of the Maxwell action in string theory \cite{Born:1933qff,Born:1934ji}
\begin{equation}
\mathcal{L}(F)=4\beta ^{2}\left( 1-\sqrt{1+\frac{F^{2}}{2\beta ^{2}}}\right)
,  \label{eq:bia}
\end{equation}%
where the parameter $\beta \sim 1/2\pi \alpha ^{\prime }$ relates to the
Regge slope. In the zero-slope limit ($\beta\to\infty $), the
action degenerates into Einstein-Maxwell theory, 
\begin{equation}
\mathcal{L}(F)=-F^{2}+\mathcal{O}\left( \frac{1}{\beta }\right) .
\label{eq:bia2}
\end{equation}
The metric of $D$-dimensional Born-Infeld AdS black hole is \cite{Cai:2004eh,Fernando:2003tz,Dey:2004yt} 
\begin{equation}
\mathrm{d}s^{2}=-f(r)\mathrm{d}t^{2}+\frac{\mathrm{d}r^{2}}{f(r)}+r^{2}\mathrm{d}\Omega _{D-2}^{2},  \label{eq:biadsm}
\end{equation}
with 
\begin{align}
\ f(r)&=1-\frac{m}{r^{D-3}}+\frac{r^{2}}{l^{2}}+\frac{4\beta ^{2}r^{2}}{
(D-1)(D-2)}\times \left( 1-\sqrt{1+\frac{(D-2)(D-3)q^{2}}{2\beta ^{2}r^{2D-4}
}}\right) \nonumber\\ 
&+\frac{2(D-2)q^{2}}{(D-1)r^{2D-6}}\times {}_{2}F_{1}\left[ \frac{D-3}{2D-4},
\frac{1}{2},\frac{3D-7}{2D-4},-\frac{(D-2)(D-3)q^{2}}{2\beta ^{2}r^{2D-4}}
\right] ,
\end{align}
where $_{2}F_{1}(a,b,c,z)$ is the hypergeometric function. The parameters $m$ and $q$ are related to the black hole mass and charge respectively as \cite{Cai:2004eh} 
\begin{align}
M =&\frac{(D-2) \Omega_{D-2}}{16\pi }m\;,\\
Q =&\sqrt{2(D-2)(D-3)}\frac{\Omega_{D-2}}{8\pi }q.
\end{align}
The event horizon $r_{+}$ is the solution of $f(r_{+})=0$. The
mass can be rewritten as 
\begin{equation}
\begin{aligned}
M=& \frac{(D-2) \Omega_{D-2}}{16 \pi} r_{+}^{D-3}\left\{1+\frac{r_{+}^{2}}{l^{2}}+\frac{4 \beta^{2} r_{+}^{2}}{(D-1)(D-2)}(1-\sqrt{1-z})\right.\\
&\left.+\frac{2(D-2) q^{2}}{(D-1) r_{+}^{2 D-6}} {}_{2}F_{1} \left[\frac{D-3}{2 D-4}, \frac{1}{2}, \frac{3 D-7}{2 D-4}, z\right]\right\},  \label{eq:m}
\end{aligned}
\end{equation}
where
\begin{equation}
z=-\frac{(D-2)(D-3)q^{2}}{2\beta ^{2}r_+^{2D-4}}.  \label{eq:z}
\end{equation}
The first law of black hole thermodynamics reads 
\begin{equation}
\mathrm{d}M=T\mathrm{d}S+\Phi \mathrm{d}Q+V\mathrm{d}P.  \label{eq:1stl}
\end{equation}
The connected Smarr relation is 
\begin{equation}
M=2(TS-VP)+\Phi Q,  \label{eq:smr}
\end{equation}
where the electric potential is
\begin{equation}
\Phi =\sqrt{\frac{D-2}{2\left( D-3\right) }}\frac{q}{r_{+}^{D-3}}{}_{2}F_{1}
\left[ \frac{D-3}{2D-4},\frac{1}{2},\frac{3D-7}{2D-4},z\right] ,
\end{equation}
and the Hawking temperature $T$, entropy $S$, thermodynamic volume $V$ are
given by \cite{Cai:2004eh,Dey:2004yt,Zou:2013owa} 
\begin{align}
T=\frac{1}{4\pi }\left[ \frac{
(D-1)r_{+}}{l^{2}}+\frac{D-3}{r_{+}}+\frac{4\beta ^{2}r_{+}}{(D-2)}\times
\left( 1-\sqrt{1-z}\right) \right],   \label{eq:biadst} 
\end{align}
\begin{align}
S=& \frac{\Omega_{D-2}}{4}r_{+}^{D-2}\;,
\;V=\frac{\Omega_{D-2}}{D-1}
r_{+}^{D-1}.  \label{eq:biadssv}
\end{align}
Substituting \cref{eq:pv,eq:cc} into \cref{eq:biadst}, one obtains the equation of
state 
\begin{equation}
P(V,T)=\frac{D-2}{4r_{+}}\left\{ T-\frac{D-3}{4\pi r_{+}}-\frac{\beta
^{2}r_{+}}{\pi (D-2)}\times \left( 1-\sqrt{1-z}\right) \right\} .
\label{eq:eos}
\end{equation}
The critical point $r_c$ obeys 
\begin{equation}
\frac{\partial P}{\partial r_{+}}|_{r=r_{c}}=\frac{\partial ^{2}P}{\partial r_{+}^{2}}|_{r=r_{c}}=0,  \label{eq:crcond}
\end{equation}
which can be exactly solved in $D=4$ case,
\begin{align}
T_{c} = & \frac{1}{2 \pi r_{c}} - \frac{Q^2}{\pi r_{c}^3} \frac{1}{\sqrt{1+ Q^2/\beta^2 r_{c}^4}}, \\
P_{c} = & \frac{1}{8\pi r_{c}^2} - \frac{Q^2}{2\pi r_{c}^4} \frac{1}{\sqrt{1+ Q^2/\beta^2 r_{c}^4}} 
- \frac{\beta^2}{4 \pi} \left( 1-\sqrt{1+\frac{ Q^{2}}{\beta^{2} r_{c}^{4}}} \right).
\end{align}

\begin{figure}[H]
\centering
\subfigure[critical point $r_c$]{\begin{minipage}[t]{0.45\textwidth}
\centering
\includegraphics[scale=0.28]{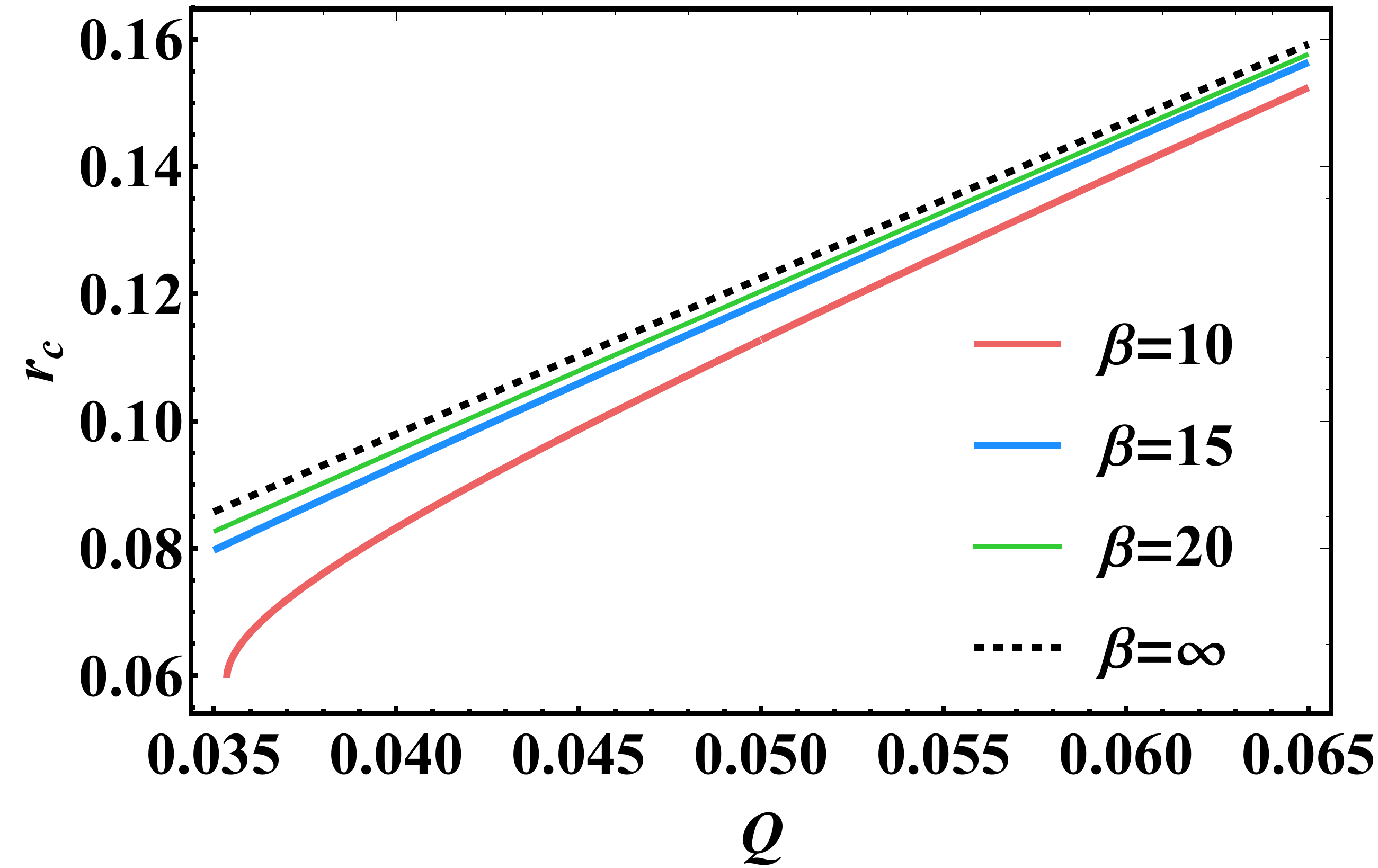}
\end{minipage}
} 
\subfigure[critical temperature $T_{c}$]{\begin{minipage}[t]{0.45\textwidth}
\centering
\includegraphics[scale=0.28]{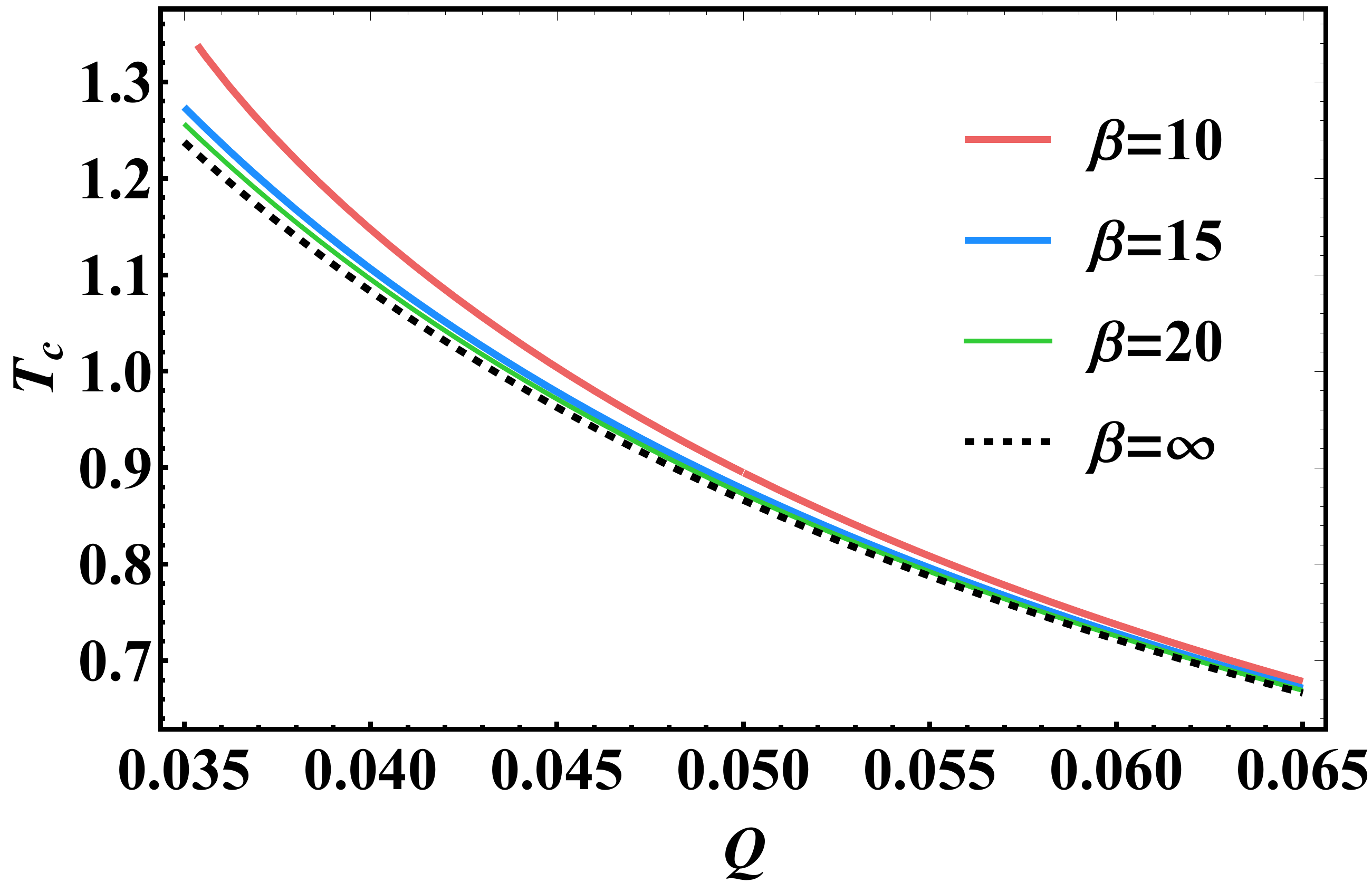}
\end{minipage}
}
\caption{The critical point $r_c$ and critical temperature $T_{c}$ versus charge $Q$ in $D=4$, from bottom to the top, the curves correspond to $\beta=$ 
10, 15, 20, $\infty$.}
\label{fig:biadsrc}
\end{figure}

We can plot the critical temperature with varying $\beta $ and $Q$ in \cref{fig:biadsrc}. We find that it is cut as $\beta Q$ is $\sqrt{2}/4$,  which is in agreement with the result of \cite{Wang:2018xdz}. Furthermore, one can easily go back to the RN-AdS case as in \cite{Okcu:2016tgt} by simply taking the limit $\beta\to\infty $.

Unlike the RN-AdS black holes\cite{Mo:2018rgq}, for $D>4$ it is very difficult to obtain an exact solution of the critical radius for Born-Infeld AdS black holes. We thus numerically solve \cref{eq:crcond}. The result shows that the critical radius is greatly suppressed by the spatial dimensionality.
\begin{figure}[H]
\centering
\includegraphics[width=10cm]{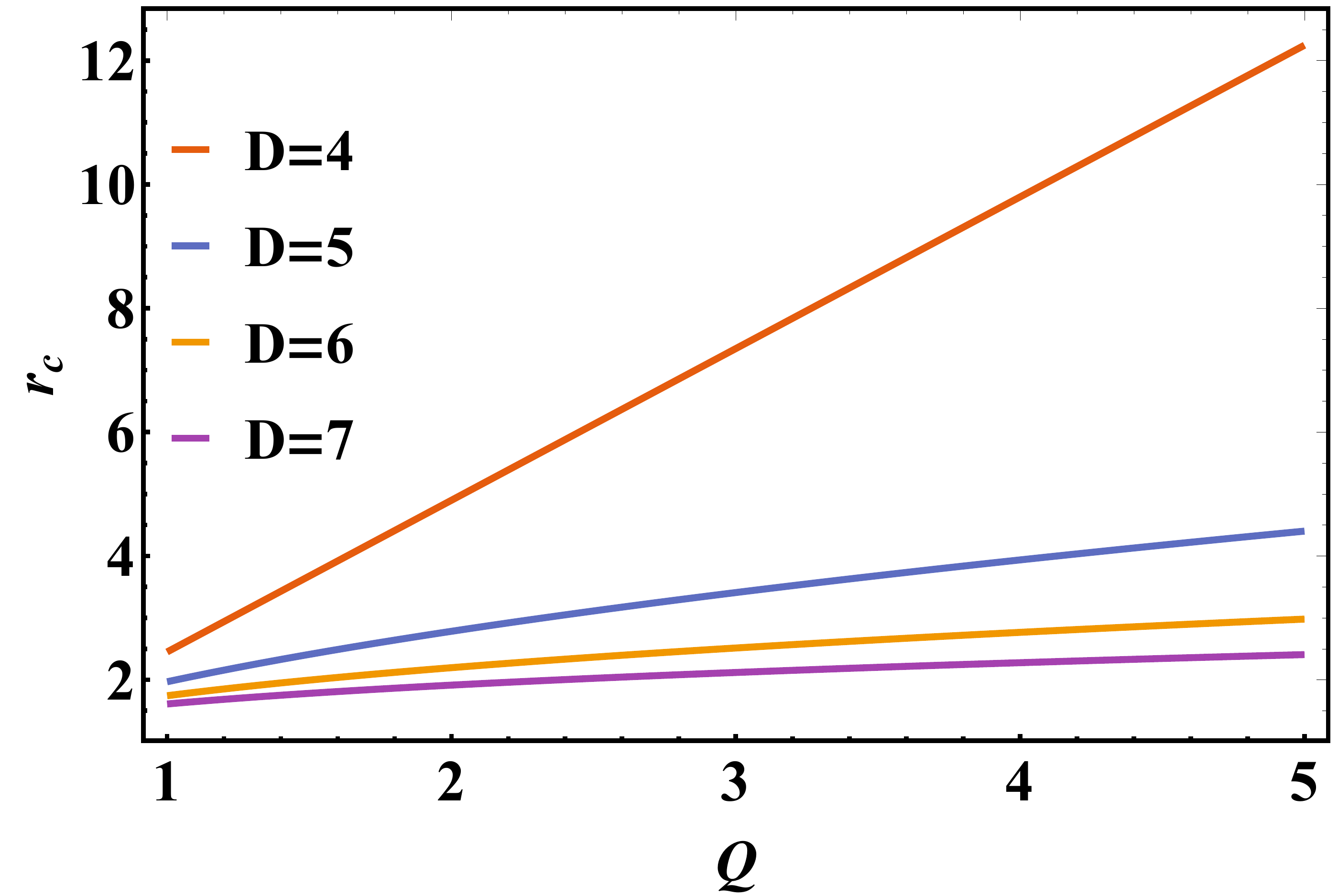}
\caption{The critical radius versus the charge $Q$ in $D=4,5,6,7$ dimensions. The solid red line is the analytic solution. In dimension higher than four, the critical radius is less sensitive to the charge. Here we take $\beta=20$.}
\label{fig:rcqD}
\end{figure}

\section{Joule-Thomson Expansion of Born-Infeld AdS Black Hole}\label{sec:jte}
In Joule-Thomson expansion, gas at a high pressure passes through a porous plug or small valve while keeping thermally insulated so that no heat is
exchanged with the environment. One can describe the temperature change by the Joule-Thomson coefficient 
\begin{equation}
\mu =\left( \frac{\partial T}{\partial P}\right) _{H}=\frac{1}{C_{P}}\left[
T\left( \frac{\partial V}{\partial T}\right) _{P}-V\right],  \label{eq:jtc}
\end{equation}
whose sign determines whether cooling or heating will occur.
And $\mu =0$ gives the inversion temperature
\begin{equation}
T_{i}=V\left( \frac{\partial T}{\partial V}\right) _{P}.  \label{eq:it0}
\end{equation}

\subsection{Van der Waals fluid}
The van der Waals equation 
\begin{equation}
P = \frac{T}{v-b} - \frac{a}{v^{2}},
\label{eq:vdw}
\end{equation}
generalizes the ideal gas equation, and is regarded as the
appropriate description of real fluids. $v=V/N$ is the specific volume
and $a,b$ measure the attraction between particles and molecule volume, respectively. The enthalpy is given by
\begin{equation}
H(T,v) = \frac{3}{2} T + \frac{Tv}{v-b} - \frac{2 a}{v}.
\label{eq:vdwH}
\end{equation}
One can spare lots of efforts to get the inversion temperature as a function of inversion pressure as follows
\begin{equation}
T_{i}=\frac{2\left( 5a-3b^{2}P_{i}\pm 4\sqrt{a^{2}-3ab^{2}P_{i}}\right) }{9b}.  \label{eq:vdwi}
\end{equation}%

\begin{figure}[H]
\centering
\includegraphics[width=10cm]{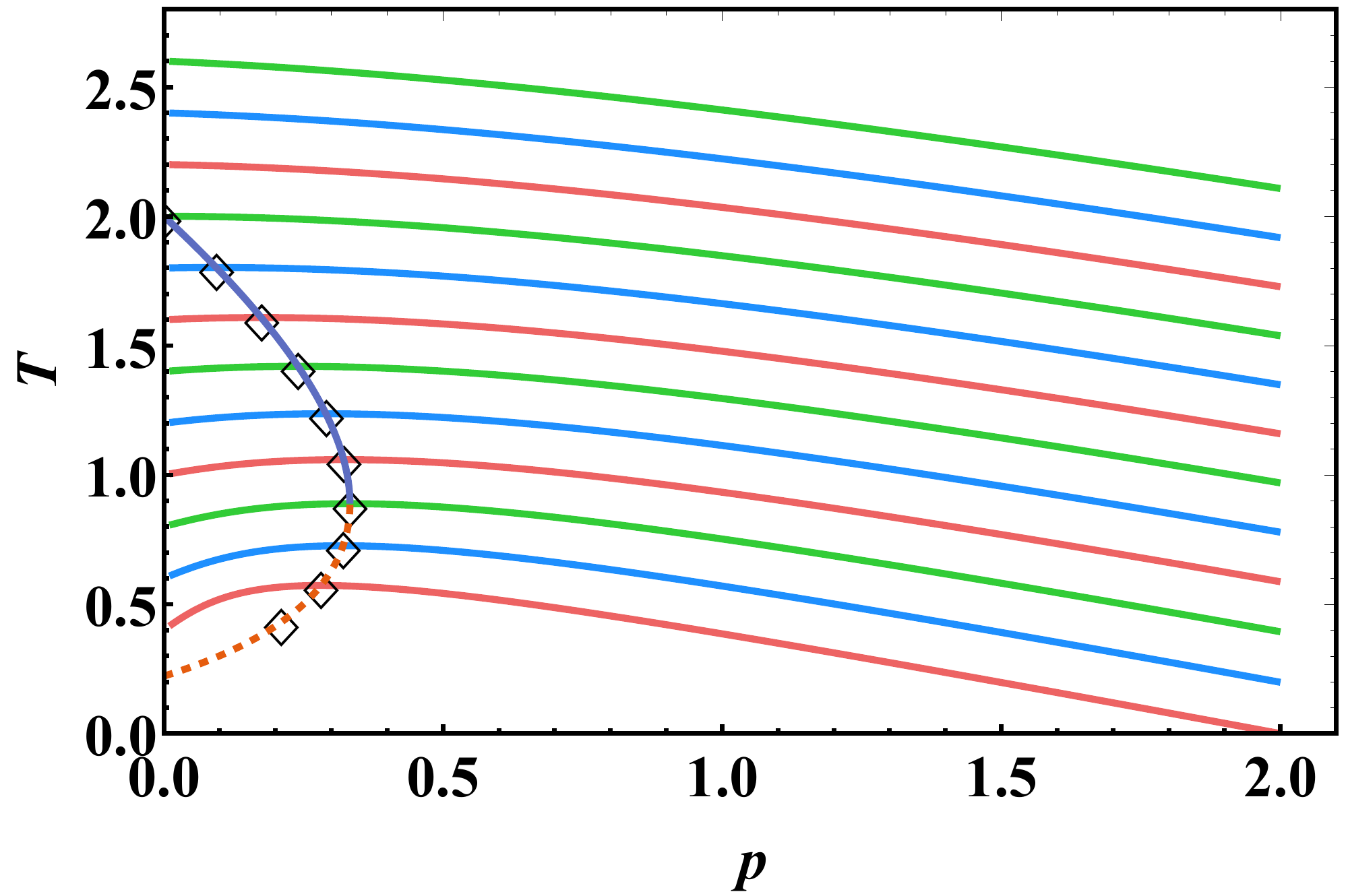}
\caption{The colored isenthalpic curves from bottom to top correspond to $H$ starting from $1$ to $6.5$ with an interval of $0.5$. Black empty diamonds pin on the maximum value points. The purple solid curve together with dashed orange one separates the cooling region and the heating
region. We have set the parameters $a=b=1$.}
\label{fig:jtvan}
\end{figure}

With \cref{eq:vdw,eq:vdwH} in hand we can plot the isenthalpic curves in the $T-P$ plane, as shown in \cref{fig:jtvan}. 
The slopes of the isenthalpic curves are positive in the cooling region and negative in the heating region, and change signs when
crossing the inversion curves.

\subsection{Born-Infeld AdS Black Hole}
Now we consider Joule-Thomson expansion of $4D$ Born-Infeld AdS Black Hole. Thermodynamic quantities can be acquired with the relations \cref{eq:1stl,eq:smr,eq:biadst,eq:biadssv}, the heat capacity at constant pressure is
\begin{equation}
C_{P} =T \left(\frac{\partial S}{\partial T}\right)_{P,Q,\beta} 
=\frac{2\pi r_{+}^{4} (8\pi Pr_{+}^{2}+1+2\beta^{2}r_{+}^{2}(1-\sqrt{1-z_4}))}{8\pi P r_{+}^{4} -r_{+}^{2}+2\beta^{2}r_{+}^{4}(1-\sqrt{1-z_4})+4Q^2/\sqrt{1-z_4}},
\end{equation}
where
\begin{equation}
z_4=z(D=4)=-\frac{Q^2}{\beta^2 r_{+}^4},
\end{equation}
and one can derive
\begin{equation}
\mu =\frac{1}{C_{P}}\left[
T\left( \frac{\partial V}{\partial T}\right) _{P,\beta}-V\right] =\frac{8 r_{+}^2+32 \pi  P r_{+}^4-8 Q^2/\sqrt{1-z_4}+8 \beta ^2 r_{+}^4 \left(1-\sqrt{1-z_4}\right)}{3r_{+} \left(1+8 \pi  P r_{+}^2+2 \beta^2 r_{+}^2 \left(1-\sqrt{1-z_4}\right)\right)}.  \label{eq:jtc2}
\end{equation}%

\begin{figure}[H]
\centering
\subfigure{\begin{minipage}[t]{0.45\textwidth}
\centering
\includegraphics[scale=0.37]{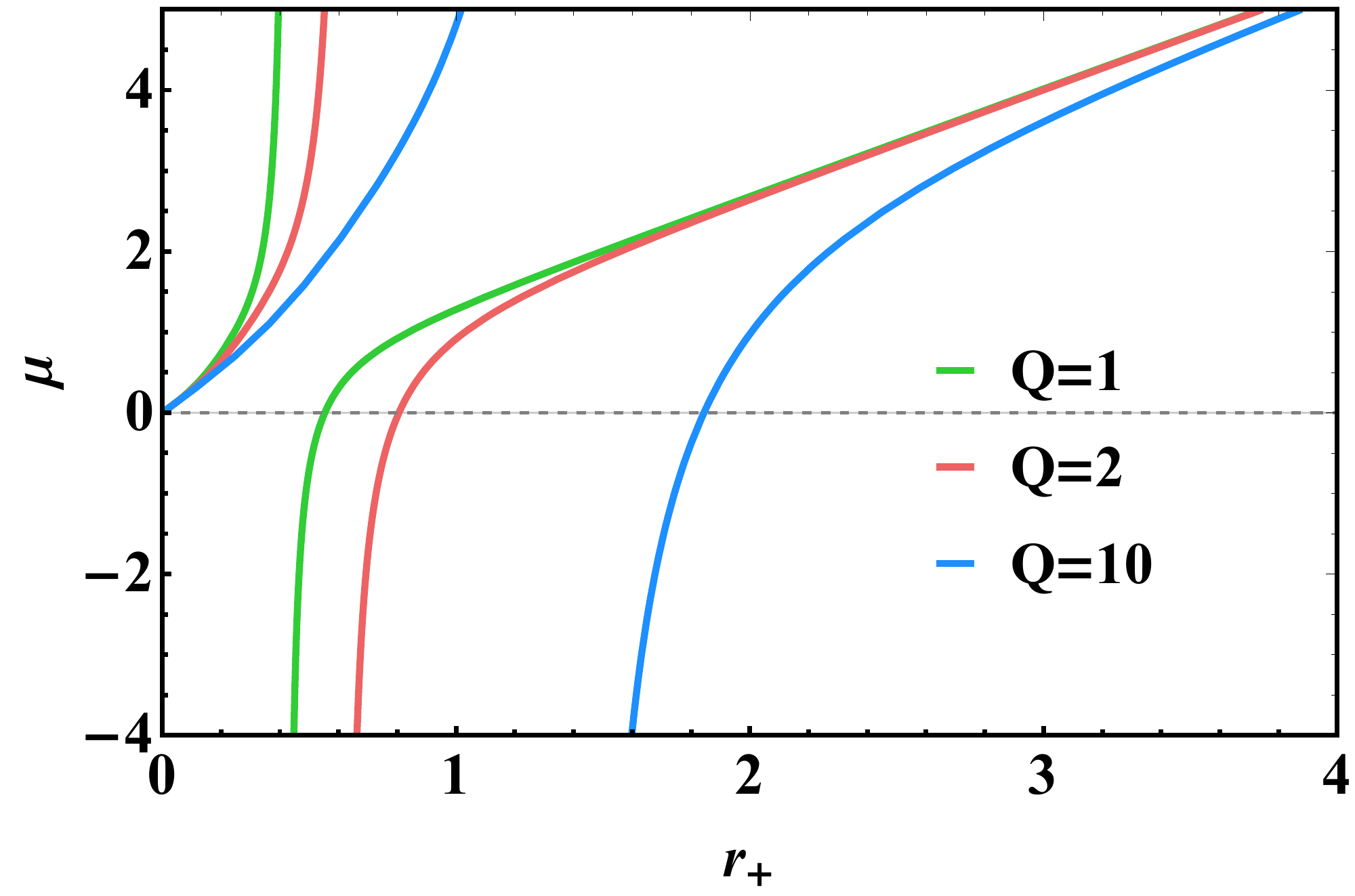}
\end{minipage}
} 
\subfigure{\begin{minipage}[t]{0.45\textwidth}
\centering
\includegraphics[scale=0.37]{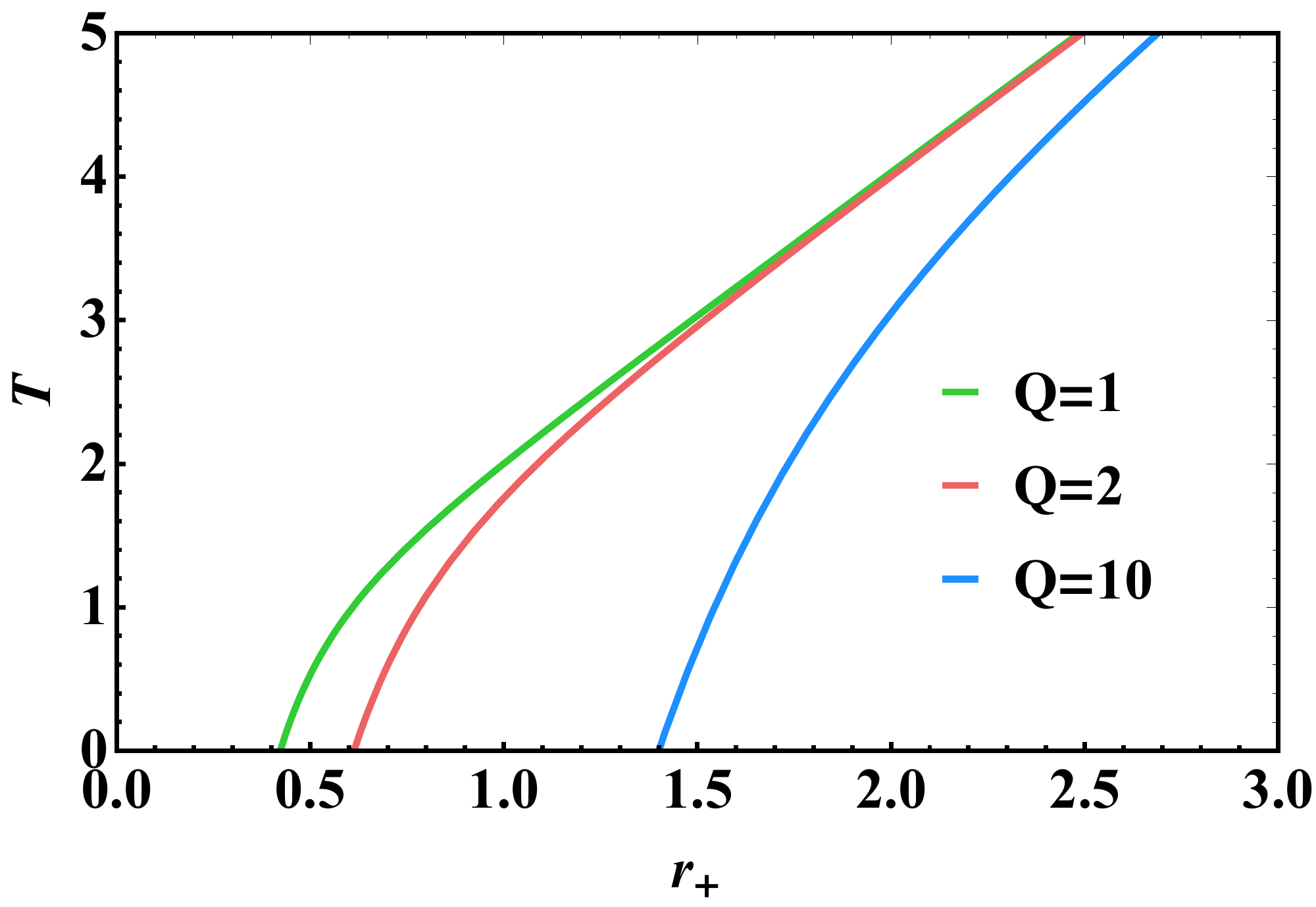}
\end{minipage}
}
\caption{Joule-Thomson coefficient $\mu$ and Hawking temperature T versus the event horizon $r_+$, here $\beta=20$, P=1, from the left to the right, the curves correspond to $Q$ = 1, 2, 10.}
\label{fig:muforbeta}
\end{figure}
The Joule-Thomson coefficient $\mu$ versus the horizon $r_+$ is shown in \cref{fig:muforbeta}. We fix $\beta=20$, the pressure $P=1$ and the charge $Q$ as 1, 2,  10 in the order. There exist both a divergence point and a zero point for different $Q$. By comparing the above two figures, we can easily see that the divergence point of the Joule-Thomson coefficient is consistent with the zero point of Hawking temperature. The divergence point here reveals the information of Hawking temperature and corresponds to the extremal black holes. 

Taking $D=4$ in \cref{eq:eos}, one obtains the equation of state 
\begin{equation}
T=2r_{+}P+\frac{1}{4\pi r_{+}}+\frac{\beta ^{2}r_{+}}{2\pi }\left( 1-\sqrt{1-z_4}\right) ,  \label{eq:eos2}
\end{equation}%
and the inversion temperature is given by 
\begin{equation}
T_{i}=V\left( \frac{\partial T}{\partial V}\right)_{P}=\frac{r_{+}}{3}%
\left\{ \frac{\beta ^{2}}{2\pi }\left( 1-\sqrt{1-z_4}\right) +\frac{Q^{2}}{\pi r_{+}^{4}\sqrt{1-z_4}}+2P-\frac{1}{4\pi r_{+}^{2}}\right\} .  \label{eq:it02}
\end{equation}
Plugging \cref{eq:it02} into \cref{eq:eos2} at $P=P_{i}$ gives 
\begin{align}
& P_i= -\frac{1}{4 \pi  r_{+}^2}-\frac{\beta^2}{4\pi}+\frac{Q^2}{2\pi  r_{+}^4 \sqrt{1-z_4}}+\frac{\beta^2}{4\pi} \sqrt{1-z_4} , \label{eq:it1} \\
& T_i= -\frac{1}{4\pi r_{+}}+\frac{Q^2}{2\pi r_{+}^3 \sqrt{1-z_4}}. \label{eq:it2}
\end{align}
\cref{eq:it1,eq:it2} make up the parameter equation of the inversion curve.
Since \cref{eq:it1,eq:it2} may have no analytical solution, we use
numerical solutions to plot inversion curves in $T-P$ plane. 
\begin{figure}[H]
\centering
\subfigure[Q=1]{\begin{minipage}[t]{0.45\textwidth}
\centering
\includegraphics[scale=0.38]{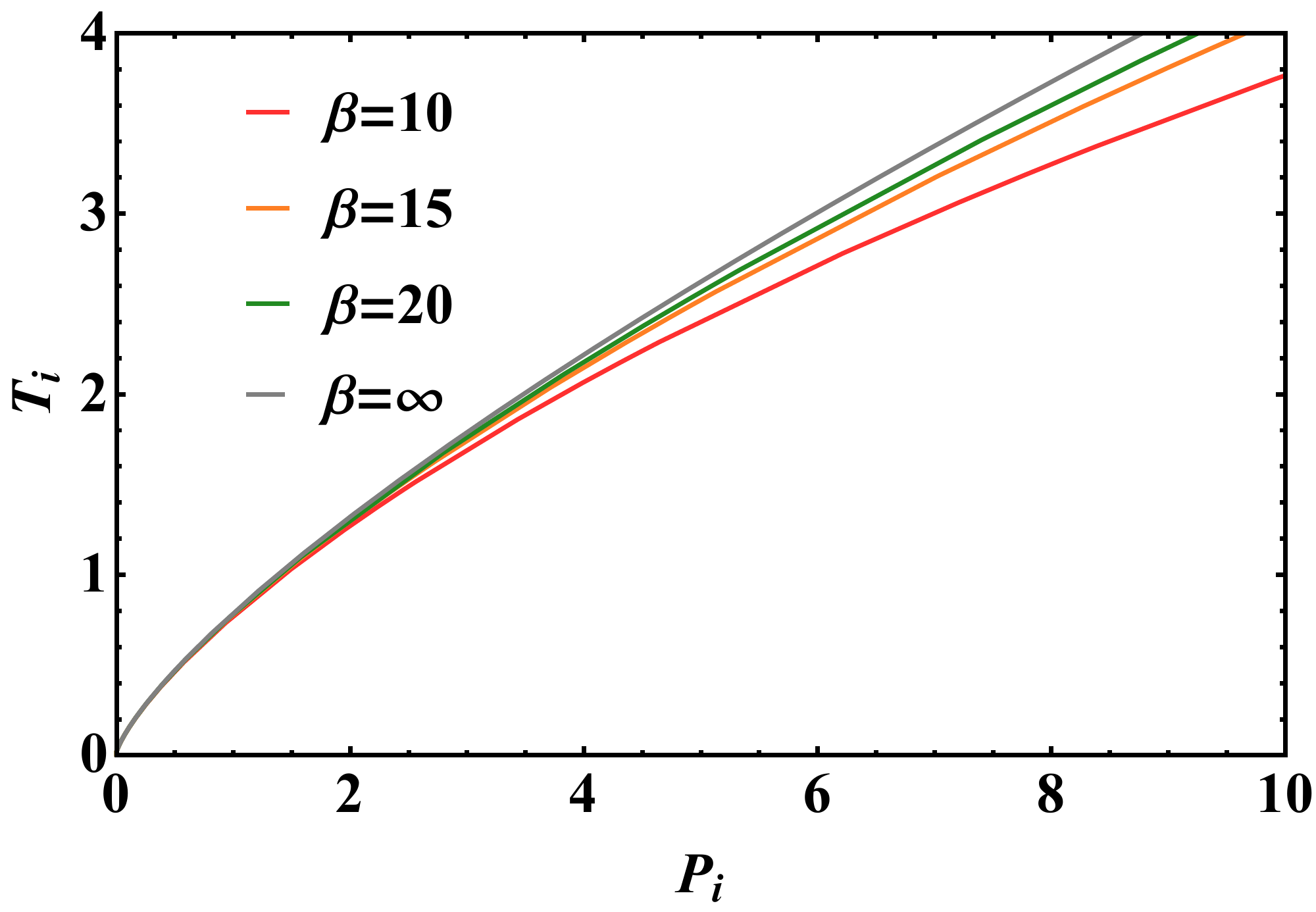}\label{fig:Tipib}
\end{minipage}
} 
\subfigure[$\beta$=20]{\begin{minipage}[t]{0.45\textwidth}
\centering
\includegraphics[scale=0.38]{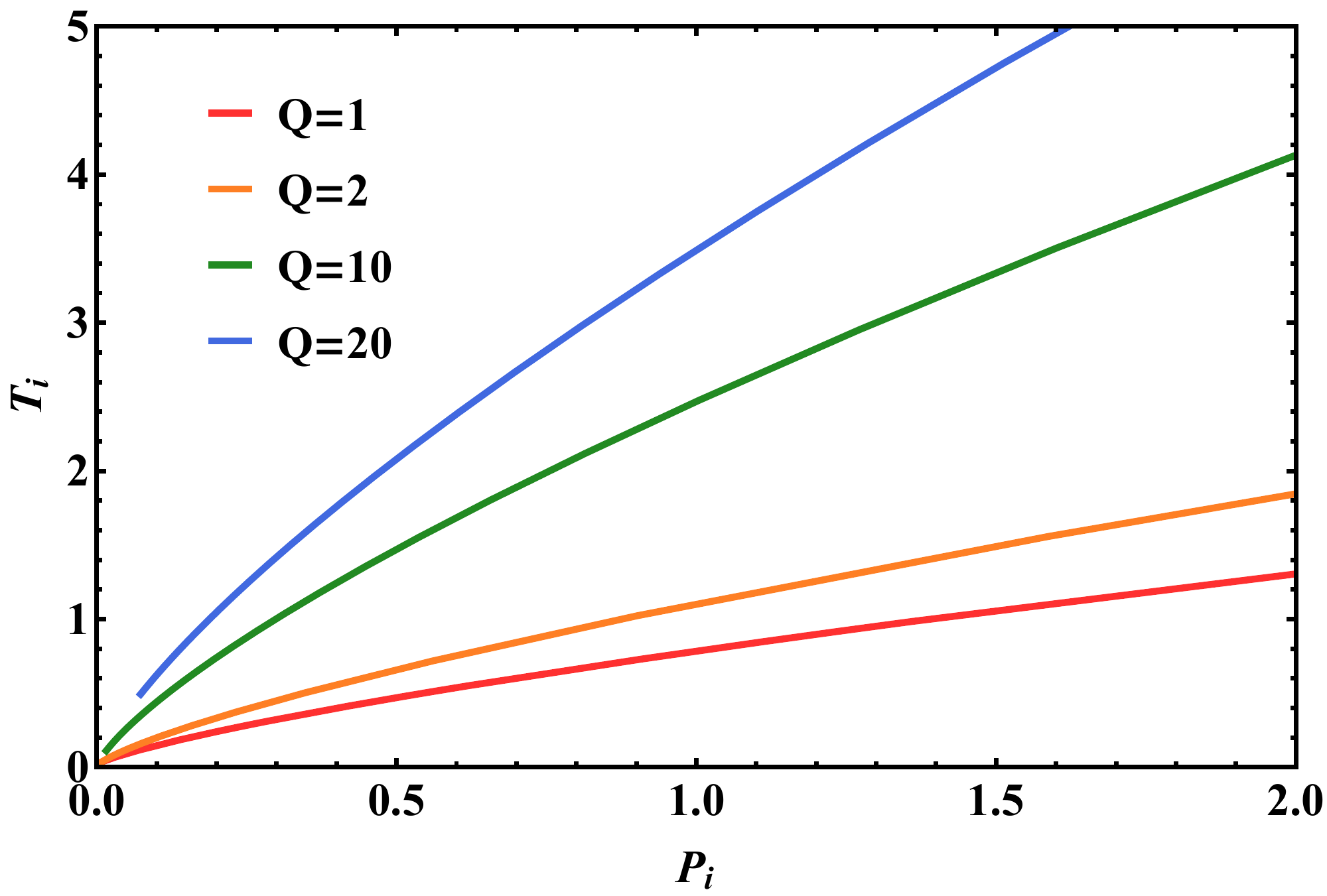}\label{fig:Tipiq}
\end{minipage}
}
\caption{Inversion curves for Born-Infeld AdS Black Holes in $T-P$ plane. From bottom to the top, the curves in the left panel correspond to $Q=1$ and $\beta$ = 10, 15, 20, $\infty $. Those in the right panel correspond to $\beta=20$ and $Q$ = 1, 2, 10, 20.}
\label{fig:Tipi}
\end{figure}

The inversion curves for different values of $\beta $ and $Q$ are shown in \cref{fig:Tipi}. The inversion temperature increases monotonically with the inversion pressure, but the slope of inversion curve decreases with the inversion pressure. Moreover, the inversion temperature increases with the charge $Q$ and parameter $\beta$. We can go back to the case of RN-AdS black hole as $\beta\to\infty$. Comparing with the van der Waals fluids, we see from \cref{fig:Tipi} that the inversion curve is not closed, and there exists only one inversion curve that corresponds to the lower dashed orange curve in \cref{fig:jtvan}. Such a difference actuates us to explore the essence of AdS black holes from the perspective of thermodynamics. The existence of inversion temperature results from the competition of attractive and repulsive interactions between real molecules. However, if we regard the AdS black hole as a thermodynamic system of certain molecules, the dominant interaction is attractive, although the electromagnetic interaction is repulsive(detail in Appendix). As a result, the Joule-Thomson coefficient is positive above the inversion curves and cooling occurs inside this region. It means that the Born-Infeld AdS black holes always cool above the inversion curve during the Joule-Thomson expansion, which is similar with the case of RN-AdS black hole \cite{Okcu:2016tgt}, Kerr-AdS black hole \cite{Okcu:2017qgo} and other AdS black holes in the previous works.

\begin{figure}[H]
\centering
\subfigure[The minimum of the  inversion temperature versus the charge $Q$]{\begin{minipage}[t]{0.45\textwidth}
\centering
\includegraphics[scale=0.36]{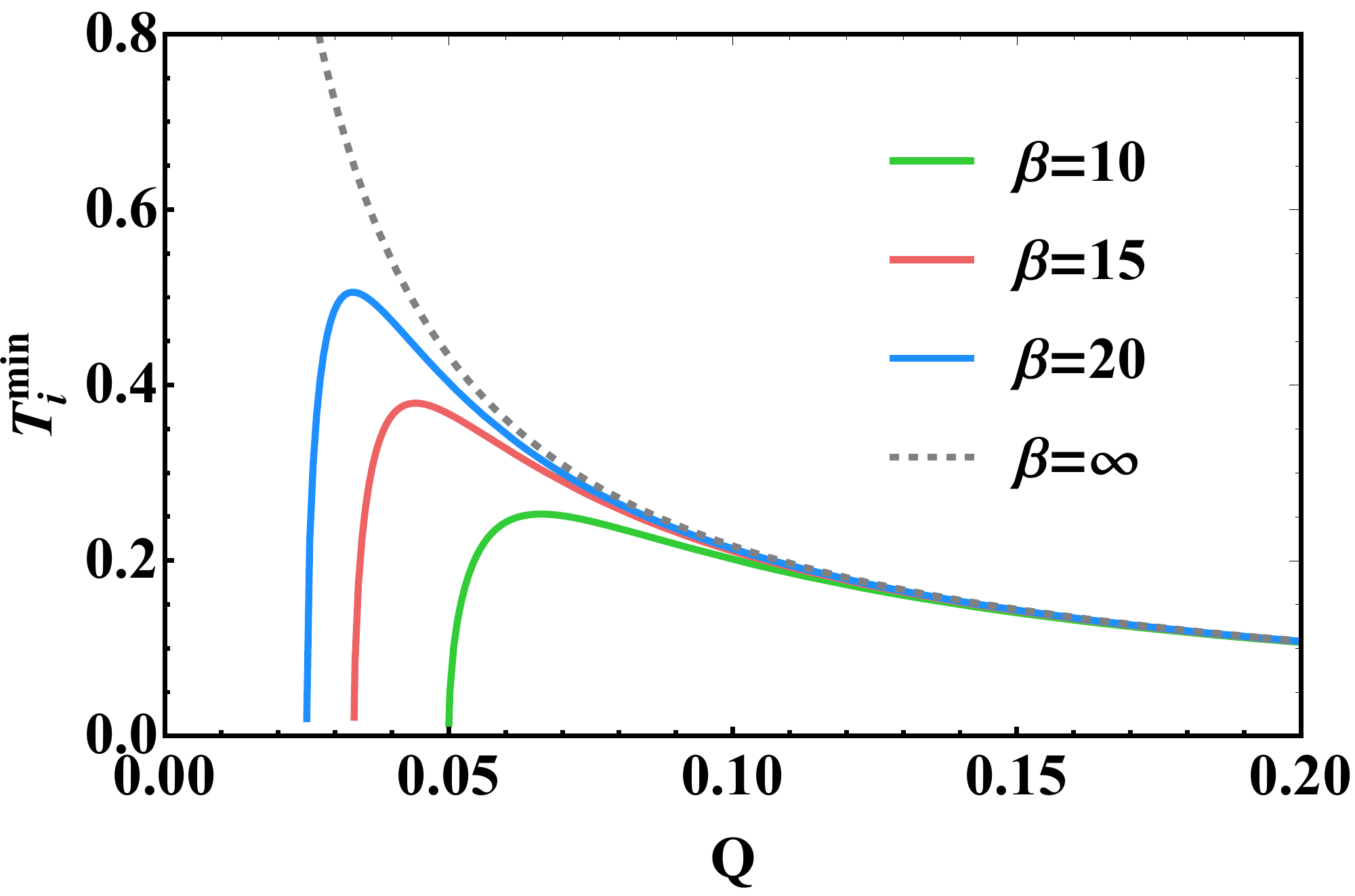}\label{fig:ratio1}
\end{minipage}
} 
\subfigure[The ratio  $T_i^{\min}/T_c$ versus the charge $Q$]{\begin{minipage}[t]{0.45\textwidth}
\centering
\includegraphics[scale=0.36]{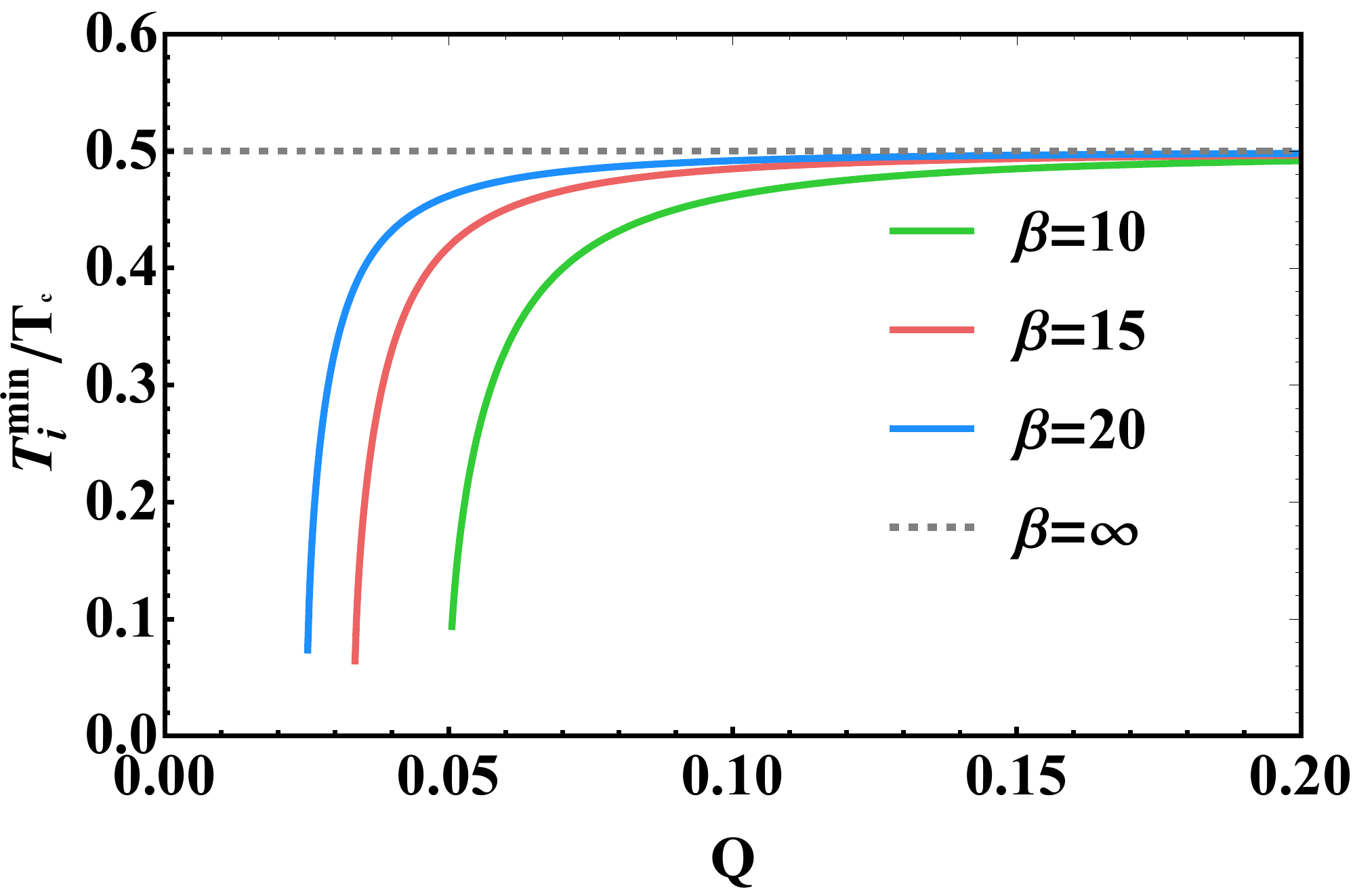}\label{fig:ratio2}
\end{minipage}
}
\caption{The minimum of the inversion temperature and the ratio  $T_i^{\min}/T_c$ versus the charge $Q$. From bottom to the top, the curves correspond to $\beta = 10, 15, 20, \infty$.}
\label{fig:ratio}
\end{figure}

Now we study the minimum of the inversion temperature, which can be obtained from \cref{eq:it2},
\begin{equation}
T_{i}^{\min } =-\frac{1}{4\pi r_{+}^{\min }}+\frac{Q^{2}}{2\pi
r_{+}^{\min 3}}\frac{1}{\sqrt{1+\frac{Q^{2}}{\beta ^{2}r_{+}^{\min 4}}}},
\end{equation}
where $r_{+}^{\min}$ is obtained by setting $ P_{i}=0$ in \cref{eq:it1}. The minimum of the inversion temperature can be obtained numerically. We then calculate the  $Q$ dependence of the  minimum of the  inversion temperature with different $\beta$ in \cref{fig:ratio1}.   We can also return to the case of RN-AdS black holes as $\beta\to\infty$. It is significant to calculate the ratio between the minimum of the inversion temperature $T_i^{\min}$ and the critical temperature $T_c$. The previous work shows that this ratio turns out to be $\dfrac{1}{2}$ \cite{Okcu:2016tgt}. 
The minimum of  the  inversion temperature and critical temperature can be obtained numerically and are displayed in \cref{fig:biadsrc} and \cref{fig:ratio1}. The $Q$ dependence of the ratio  $T_i^{\min}/T_c$  with different $\beta$ is shown in \cref{fig:ratio2}.  We can see that the ratio is not always $\dfrac{1}{2}$. The curves show that the ratio tends to be $\dfrac{1}{2}$ as Q increases.  When $\beta\to\infty$, it degenerates into RN-AdS black holes again and the ratio is always $\dfrac{1}{2}$. 

Other thermodynamic quantities can be calculated with the relation \cref{eq:1stl,eq:smr}. Using \cref{eq:m}, we can get the mass of this black hole in terms of $r_{+}$ in $D=4$, 
\begin{equation}
M=\frac{r_{+}}{2}\left\{ 1+\frac{8\pi P r_{+}^{2}}{3}+\frac{2\beta ^{2}r_{+}^{2}}{3}\left( 1-\sqrt{1-z_4}\right) +\frac{4Q^{2}}{3r_{+}^{2}}  {}_{2}F_{1}\left[ \frac{1}{4},\frac{1}{2},\frac{5}{4},z \right] \right\} .  \label{eq:Mr}
\end{equation}%

\begin{figure}[H]
\centering
\subfigure[$Q=1,\beta=10,M = 1.5, 1.7, 1.9, 2.1, 2.3, 2.5, 2.7$]{
\begin{minipage}[t]{0.45\textwidth}
\centering
\includegraphics[width=7.6cm]{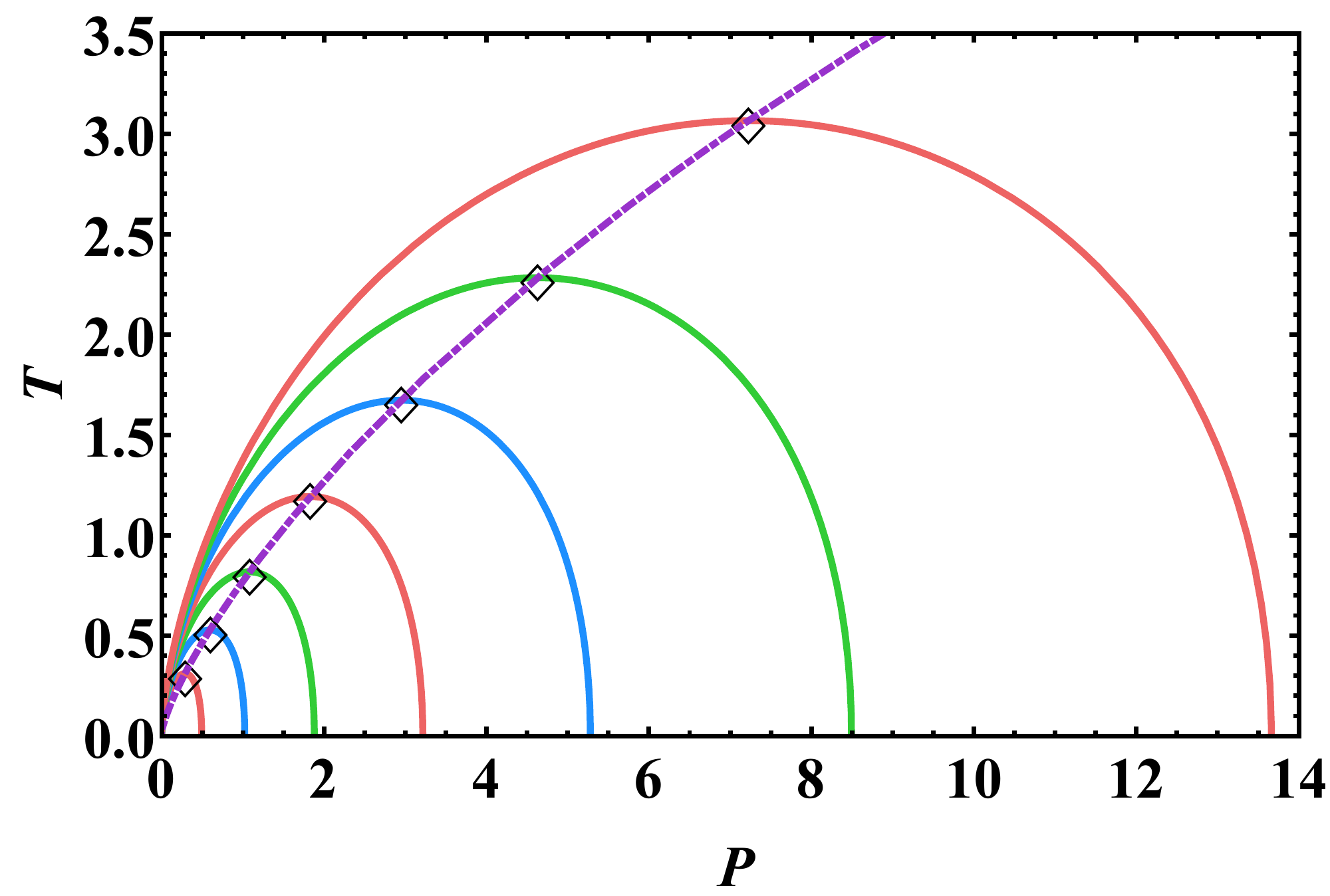}
\end{minipage}
}
\subfigure[$Q=1,\beta=20,M = 1.5, 1.7, 1.9, 2.1, 2.3, 2.5, 2.7$]{
\begin{minipage}[t]{0.45\textwidth}
\centering
\includegraphics[width=7.6cm]{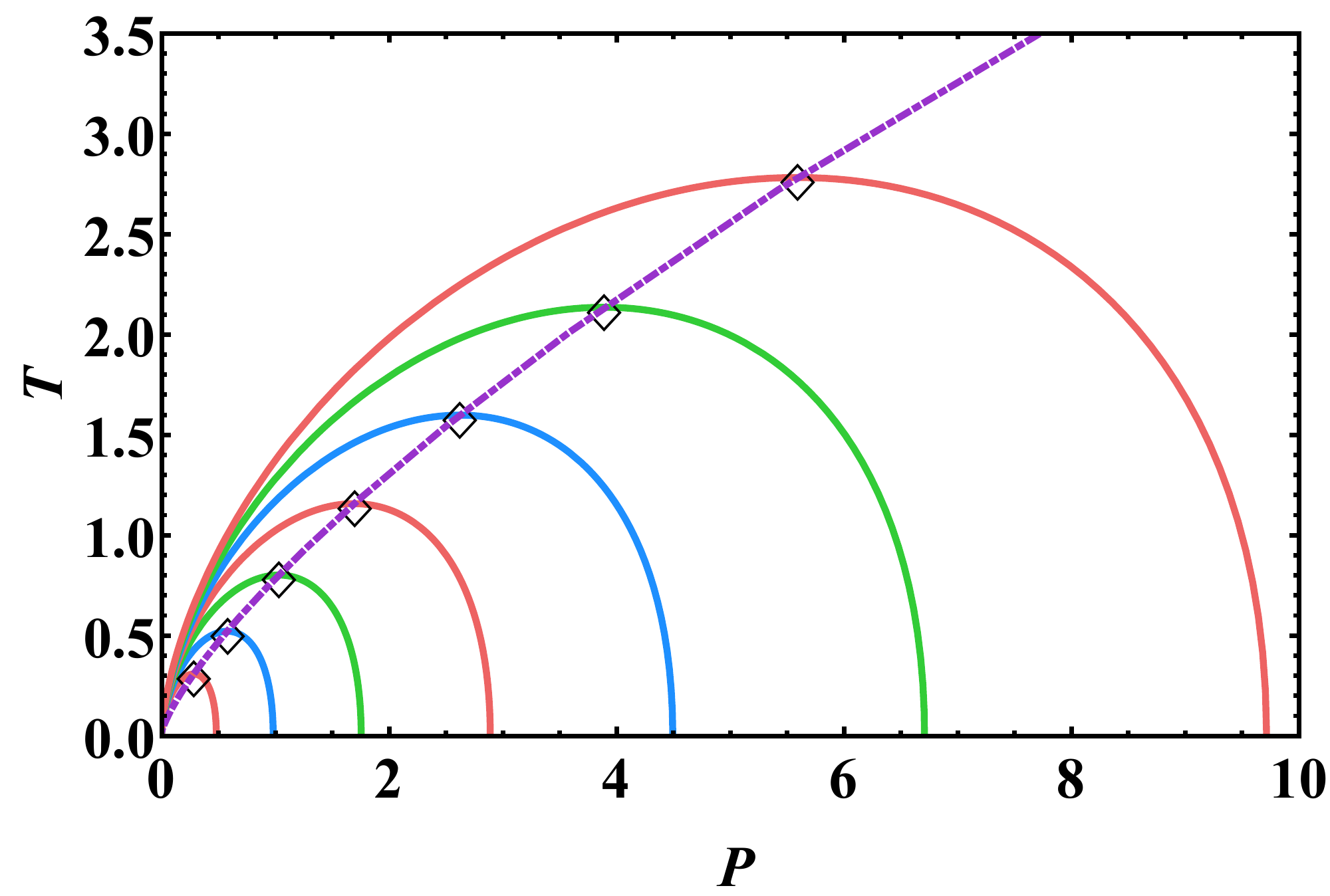}
\end{minipage}
}
\quad
\subfigure[$Q=2,\beta=10,M = 2.5, 2.7, 2.9, 3.1, 3.3, 3.5, 3.7$]{
\begin{minipage}[t]{0.45\textwidth}
\centering
\includegraphics[width=8cm]{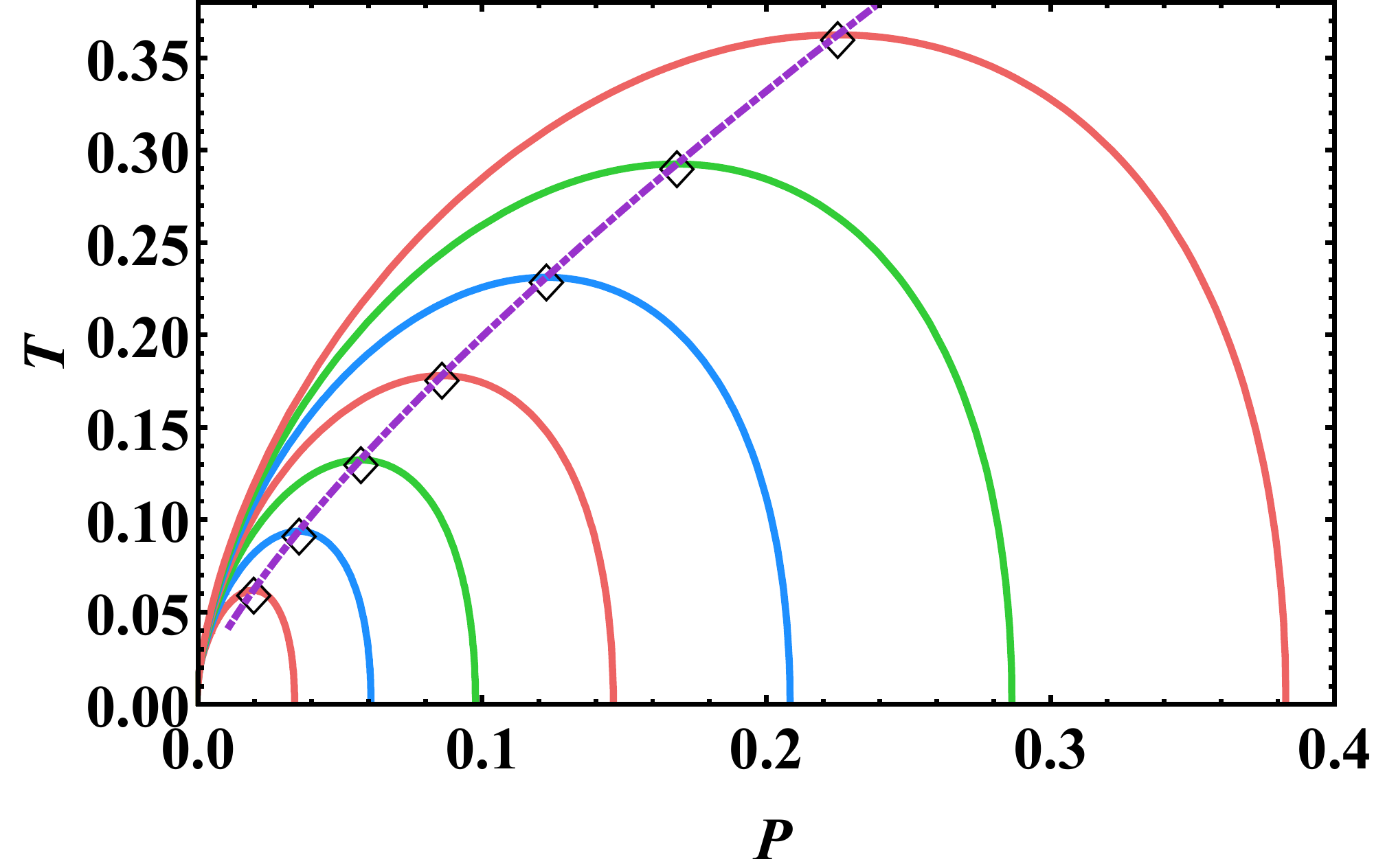}
\end{minipage}
}
\quad
\subfigure[$Q=1,\beta=\infty,M = 1.5, 1.7, 1.9, 2.1, 2.3, 2.5, 2.7$]{
\begin{minipage}[t]{0.45\textwidth}
\centering
\includegraphics[width=7.6cm]{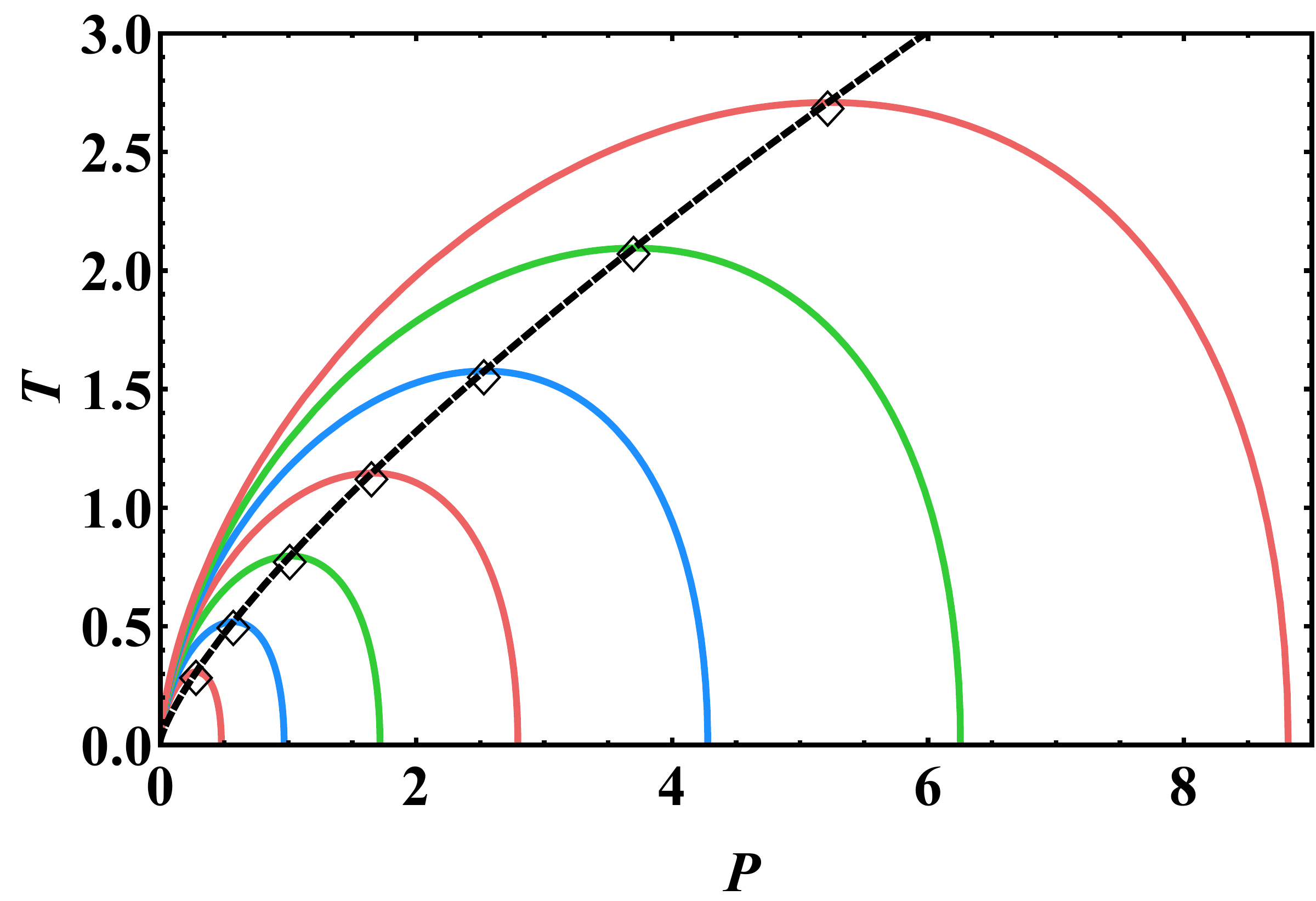}
\end{minipage}
}
\caption{The different solid lines are isenthalpic curves with parameters given below the figures. 
Black empty diamonds pin on the maximum value points, and purple dotdashed lines are the inversion curves. 
The lower right panel denotes the RN-AdS case.}
\label{fig:jtbiads}
\end{figure}

Since the Joule-Thomson expansion is an isenthalpic process, it is significant to study the isenthalpic curves of Born-Infeld AdS black holes. In the extended phase space, the mass could be interpreted as enthalpy.  So we can plot isenthalpic curves in $T-P$ plane by fixing the mass of the black hole. The isenthalpic curves are given by \cref{eq:eos2} with $r_{+}$ being the larger root of \cref{eq:Mr} at given $M$. We show the isenthalpic curves and the inversion curves of Born-Infeld AdS black holes in \cref{fig:jtbiads}, and this result is consistent with that in \cref{fig:Tipi}. The inversion curve is the dividing line between heating and cooling. The isenthalpic curve has a positive slope above the inversion curve. So there is cooling above this inversion curve. On the contrary, the sign of the slope changes and heating occurs below the inversion curve. 

Finally, we make a brief extension of the discussion above to arbitrary dimension $D>4$. The critical pressure and temperature can be obtained from \cref{eq:crcond}
\begin{equation}\bea
P_{c} = & \frac{(D-2)(D-3)}{16\pi r_{c}^2} + \frac{\beta^2 (D-2) z}{4 \pi \sqrt{1-z}} - \frac{\beta^2 \left(1-\sqrt{1-z}\right)}{4 \pi},\\
T_{c} = & \frac{D-3}{2\pi r_{c}} + \frac{\beta^2 z r_{c}}{\pi \sqrt{1-z}}.
\eea\end{equation} 
One should not forget that $z$ is the function of $r_{+}$, and in the equations above should take $r_{+}=r_{c}$.
However, the critical radius may have no analytical solution and we have to sort to numerical method. The results are shown in \cref{fig:rcqD}.
On the other hand, the inversion pressure and temperature are
\begin{equation}\bea
P_{i} = &-\frac{D(D-3)}{16\pi r_{+}^2} - \frac{\beta^2 z}{4\pi \sqrt{1-z}} - \frac{\beta^2 \left(1-\sqrt{1-z}\right)}{4 \pi},\\
T_{i} = &-\frac{D-3}{2(D-2)\pi r_{+}} - \frac{\beta^2 z r_{+}}{(D-2)\pi \sqrt{1-z}}.
\eea\end{equation} 
The above two equations define the parameter equations of the inversion curve $T(P)$. We then numerically study the 
inversion curves in various dimensions and the result is presented in \cref{fig:icd45678}:
\begin{figure}[H]
\centering
\includegraphics[width=10cm]{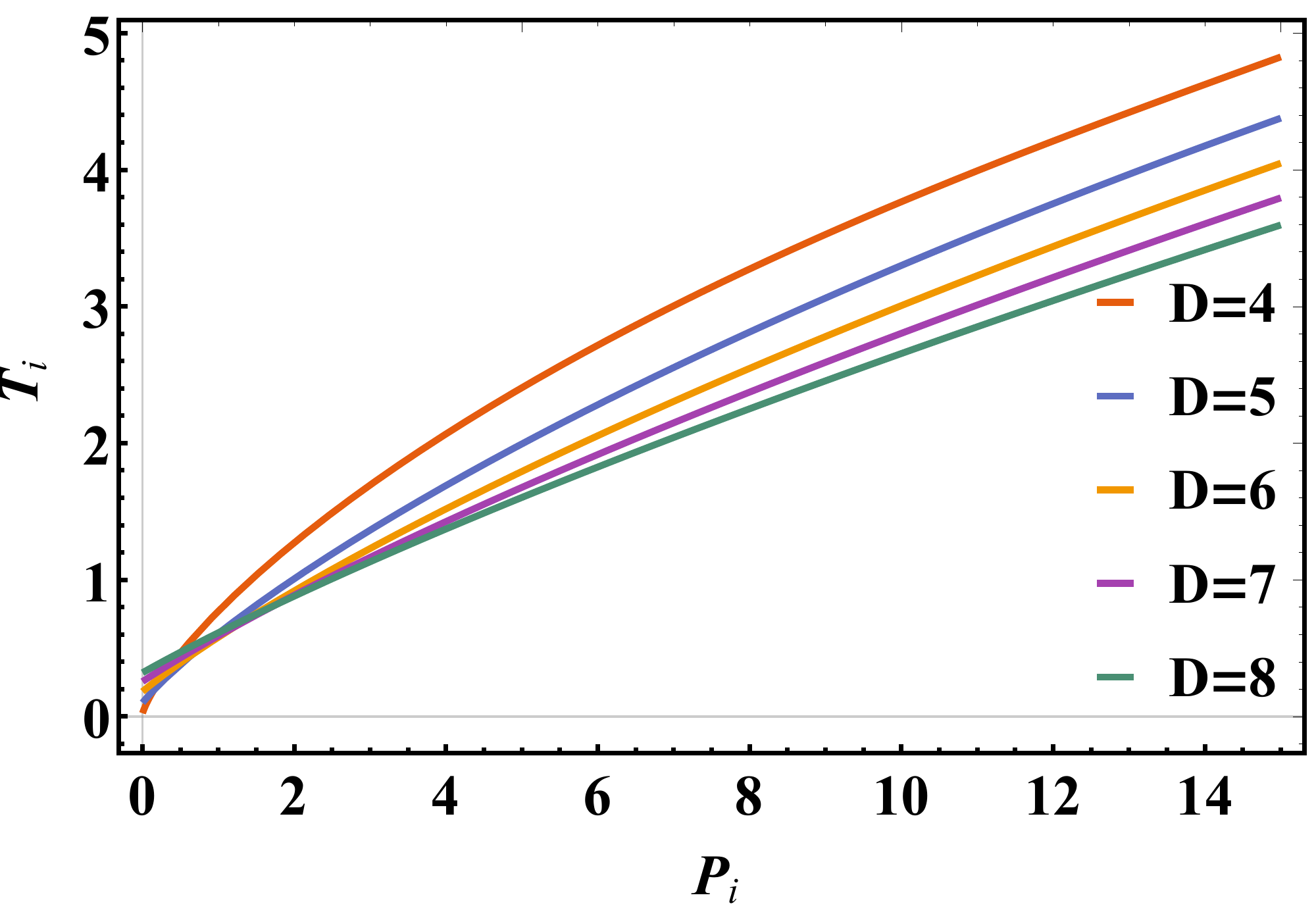}
\caption{Inversion curves in $D=4,5,6,7,8$  from top to bottom. We set $Q=1$ and $\beta=10$.}
\label{fig:icd45678}
\end{figure}
Similar effect of dimensionality is demonstrated. At low pressure, the inversion temperature has a small increase with the dimension $D$, while at high pressure it is suppressed.

To confirm our result we further show the isenthalpic curves in five and six dimensional spacetimes, respectively. 
And as shown below the inversion curves pass through the maximum points and separate the cooling and heating regions.
\begin{figure}[H]
\centering
\subfigure[$Q=1,\beta=10,M = 1.5, 1.7, 1.9, 2.1, 2.3, 2.5, 2.7$]{
\begin{minipage}[t]{0.45\textwidth}
\centering
\includegraphics[width=7.6cm]{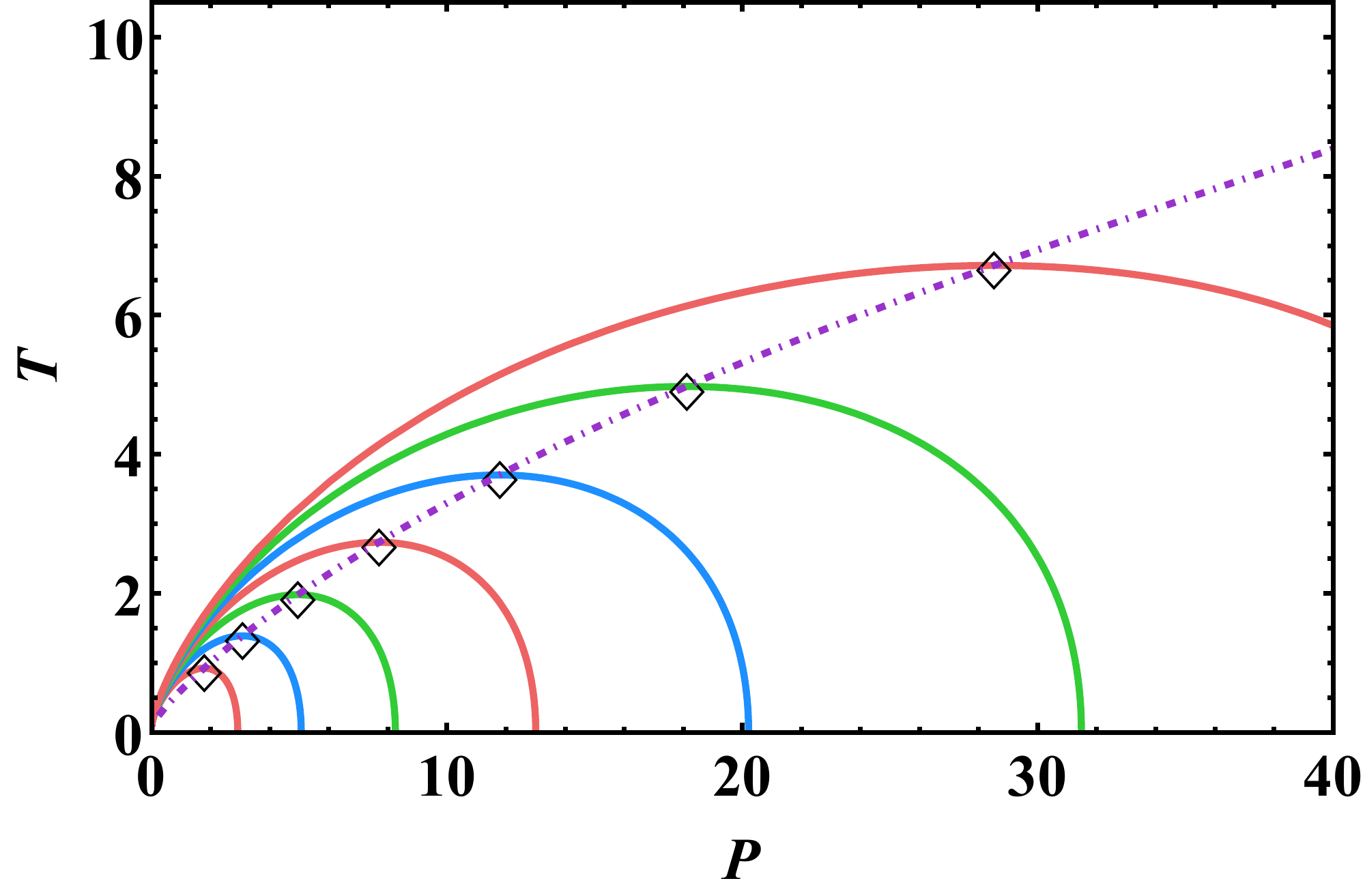}
\end{minipage}
}
\subfigure[$Q=1,\beta=10,M = 1.5, 1.7, 1.9, 2.1, 2.3, 2.5, 2.7$]{
\begin{minipage}[t]{0.45\textwidth}
\centering
\includegraphics[width=7.6cm]{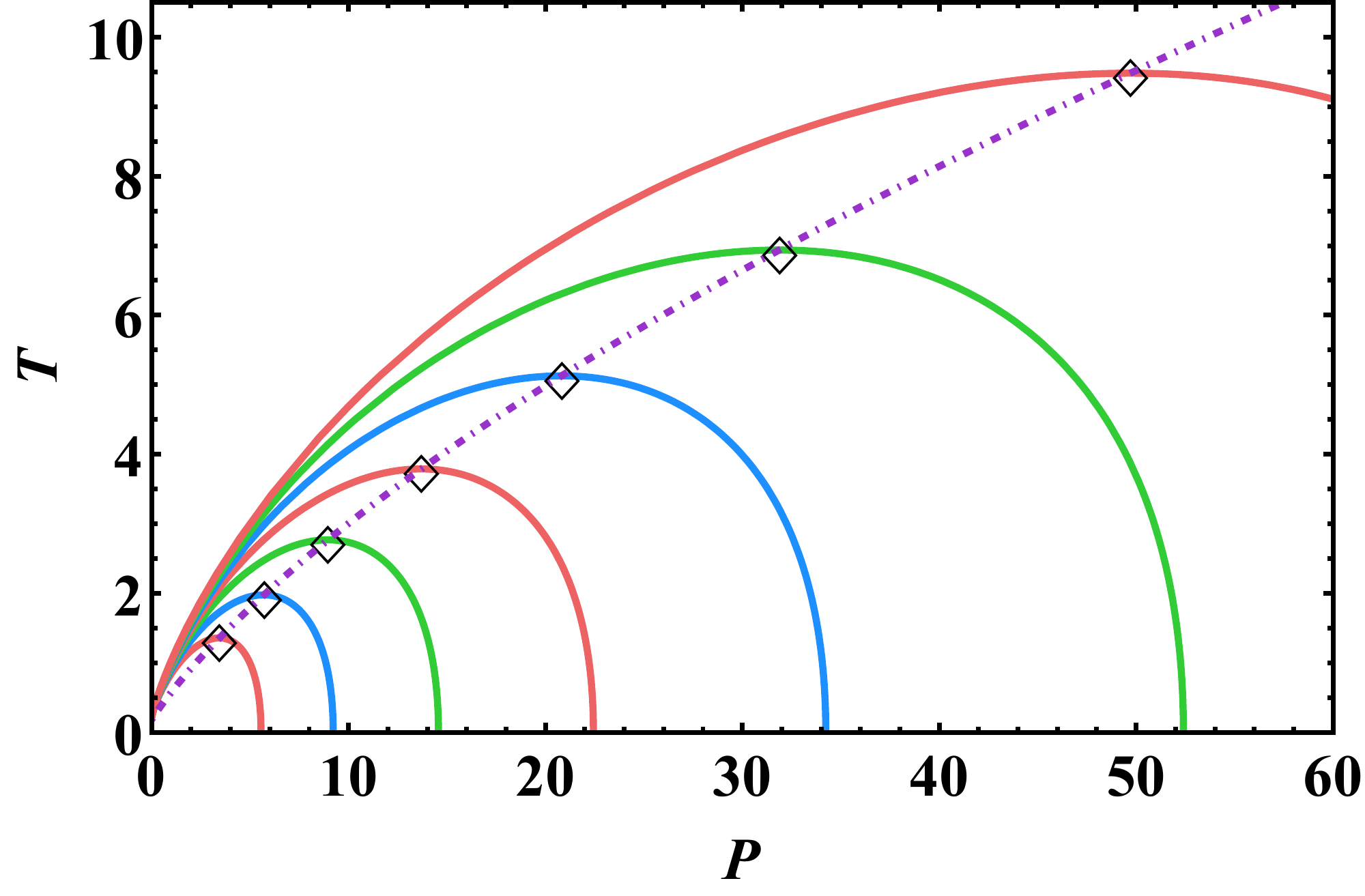}
\end{minipage}
}
\caption{The red, blue, and green solid lines are isenthalpic curves in (a) D=5 and (b) D=6 dimensions. 
The black square denote the maximum of the isenthalpic lines and the purple dashed lines represent the inversion curves.}
\label{fig:jtbiadsd56}
\end{figure}
The ratio between $T_{i}^{\min}$ and $T_{c}$ is also numerically calculated. In the previous work \cite{Mo:2018rgq}, 
the ratios for RN-AdS black holes in arbitrary dimension have been obtained. We make the comparison between our results and those in RN-AdS black holes. The dashed lines are the ratios for Born-Infeld black holes and solid lines are the cases 
of RN-AdS black holes. When $Q$ increases the ratios asymptotically reach the predicted results in \cite{Mo:2018rgq}.
However, string effects lead to obvious difference between Born-Infeld and RN-AdS cases in the small Q regime.
\begin{figure}[H]
\centering
\includegraphics[width=10cm]{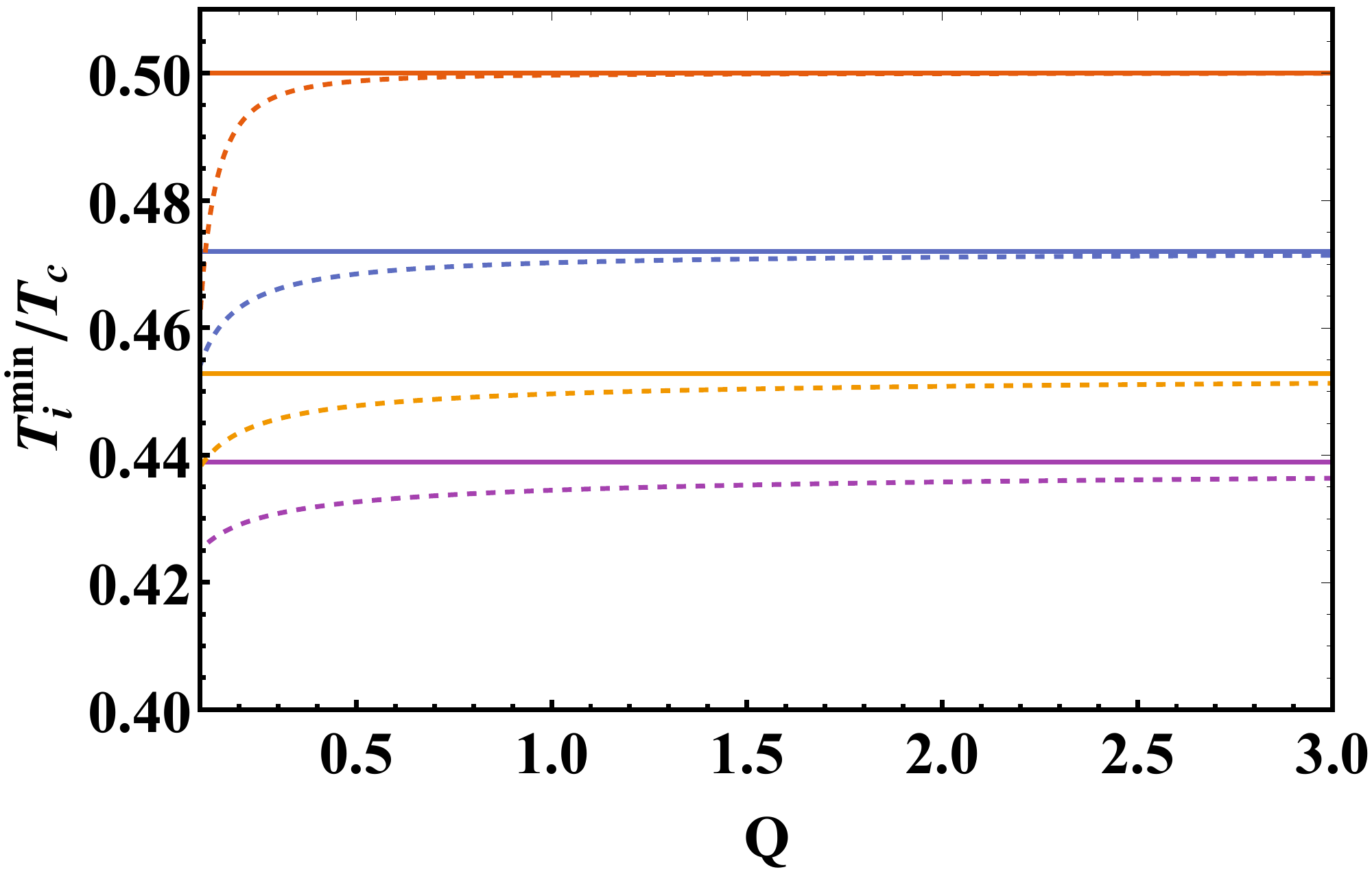}
\caption{The ratio $T_i^{\min}/T_c$ versus the charge $Q$. We set $\beta=10$.
From top to bottom the dashed lines and its asymptotic lines correspond to $D=4,5,6,7$.}
\label{fig:ratioqD}
\end{figure}

\section{Conclusion}\label{sec:con}
In this paper, we have studied the Joule-Thomson expansion for Born-Infeld AdS black holes in the extended phase space, where the cosmological constant is identified with the pressure. 
We mainly focused on $D=4$ dimension and extensions to higher dimensions were also presented.
Since the black hole mass is interpreted as enthalpy, it is the mass that does not change during the Joule-Thomson expansion. 
The Joule-Thomson coefficient $\mu$ versus the horizon $r_+$ is shown in \cref{fig:muforbeta}. 
There exist both a divergence point and a zero point in different $\beta$. We can easily see that the divergence point of the Joule-Thomson coefficient 
is consistent with the zero point of Hawking temperature which corresponds to the extremal black holes. 
In addition, one can easily go back to the RN-AdS case as in \cite{Okcu:2016tgt,Mo:2018rgq} by taking the limit $\beta\to\infty$.

The inversion curves depending on the charge $Q$ and parameter $\beta$ were investigated in Born-Infeld AdS black holes. The results were depicted in \cref{fig:Tipi}.  We also showed the isenthalpic curves and the inversion curves in \cref{fig:jtbiads}. 
Such results were reproduced in higher dimensions as demonstrated in \cref{fig:icd45678,fig:jtbiadsd56,fig:ratioqD}.
It shows that the inversion curve always has positive slope, which is consistent with previous work \cite{Okcu:2016tgt,Mo:2018rgq}. 
This means that Born-Infeld AdS black holes always cool above the inversion curve during the expansion. 
We can use the inversion curve to distinguish the cooling and heating regions for different values of $\beta $ and $Q$. 
Furthermore, we checked the ratio of the critical temperature and the minimum of the inversion temperature in \cref{fig:ratio}, 
which shows that the ratio is asymptotically $1/2$ as $Q$ increases or $\beta\to\infty$ in $D=4$, 
and those limit values of RN-AdS case in arbitrary dimension.

Finally, we would like to stress that in the present work we only focus on research of Joule-Thomson expansion for Born-Infeld AdS black holes. These results are related to many other interesting problems which deserve future study. For example, the nontrivial charged black hole solutions can be obtained within the nonlinear massive gravity theory \cite{Cai:2013prd} as well, and the black holes may present more structures when higher order corrections have been taken into account, such as asymptotically safe gravity\cite{Cai:2010jcap}, or quadratic gravity\cite{Cai:2016jhep}. Then, it becomes curious if the Joule-Thomson expansion can be applied into these theoretical paradigms and may inspire the follow-up works. However, we leave such a study for future investigation.

\section*{Acknowledgment}

We are grateful to thank Bo Ning, Peng Wang and Wei Hong for useful discussions. 
 This work is supported by NSFC (Grant No.11947408).

\section*{Appendix}

In this appendix, we make a quantitative discussion on the non-existence of maximum inversion temperature for Born-Infeld AdS black holes. To make it concrete, let us compare the van der Waals fluids and Born-Infeld AdS black holes. First, we take the Virial expansion for the state equation of van der Waals fluid to the second order\cite{Landau:vol5}:
\begin{equation}
\frac{PV}{T}=1+\frac{1}{V}B_{2}(T),
\end{equation}
where the second Virial coefficient ${B_2}(T)= b-\dfrac{a}{T}$. The modification term $\dfrac{1}{V}{{B}_{2}}(T) \ll 1$ 
and we can bring the zero-order approximation $\dfrac{1}{V}=\dfrac{P}{T}$(high temperature and low pressure limit), 
and the modified volume is $V=\dfrac{T}{P}+{{B}_{2}}(T)$. Bringing the expression above into the Joule-Thomson coefficient and we get
\begin{equation}
\mu =\frac{1}{C_{P}}\left( T\frac{\mathrm{d}{{B}_{2}}}{\mathrm{d}T}-{{B}_{2}} \right)=\frac{1}{C_{P}}\left( \frac{2a}{T}-b \right).
\end{equation}
At low temperature, the attractive interaction is dominant and ${B}_{2}$ is negative, we thus have $\mu>0$. 
When the temperature is high enough, the repulsive interaction is dominant and ${B}_{2}$ is positive. 
So $\mu$ may be less than zero. The competition of attractive and repulsive interaction results in the inversion temperature. 
Such a simple argument also gives the right $T_{i}^{\max}=2a/b$. The analysis is suitable in high temperature and low-pressure limit. 

Now we turn to the Born-Infeld AdS black holes. The equation of state is
\begin{equation}
P(V,T)=\frac{D-2}{4r_{+}}\left\{ T-\frac{D-3}{4\pi r_{+}}-\frac{\beta
	^{2}r_{+}}{\pi (D-2)}\times \left( 1-\sqrt{1-z}\right) \right\} .
\end{equation}
We see the second Virial coefficient ${{B}_{2}}(T) \sim -\dfrac{D-3}{4\pi T}$ is always negative in an arbitrary dimension, 
which means that the attractive interaction is dominant. Thus, the Joule-Thomson coefficient is approximately 
\begin{equation}
\mu =\frac{1}{C_{P}}\left( T\frac{\mathrm{d}{{B}_{2}}}{\mathrm{d}T}-{{B}_{2}} \right)=\frac{1}{C_{P}}\times \frac{1}{2\pi T},
\end{equation} 
We see that in the high temperature and low-pressure limit, the Joule-Thomson coefficient is always positive, there is no maximum inversion temperature.


\end{document}